\documentclass[10pt]{bmc_article}

\usepackage{cite} 
\usepackage{url}  
\usepackage{ifthen}  
\usepackage{multicol}   
\usepackage[utf8]{inputenc} 
\usepackage{epsfig}
\usepackage{url}

\usepackage{soul}

\newcommand{\be}{\begin{equation}}
\newcommand{\ee}{\end{equation}}
\newcommand{\bc}{\begin{center}}
\newcommand{\ec}{\end{center}}
\newcommand{\ba}{\begin{eqnarray}}
\newcommand{\ea}{\end{eqnarray}}
\newcommand{\ie}{{\it i.e.}}

\newboolean{publ}

\newenvironment{bmcformat}{\begin{raggedright}\baselineskip20pt\sloppy\setboolean{publ}{false}}{\end{raggedright}\baselineskip20pt\sloppy}

\begin{document}
\begin{bmcformat}


  \title{Scaling properties of protein family phylogenies}


\author{E. Alejandro Herrada\correspondingauthor$^1$%
       \email{E. Alejandro Herrada\correspondingauthor - alejandro@ifisc.uib-csic.es}%
      and
         V\'ictor M. Egu\'iluz$^1$%
         \email{V\'ictor M. Egu\'iluz - victor@ifisc.uib-csic.es}
       and
         Emilio Hern\'andez-Garc\'ia$^1$%
         \email{Emilio Hern\'andez-Garc\'ia - emilio@ifisc.uib-csic.es}
      and
         Carlos M. Duarte$^{2,3}$%
         \email{Carlos M. Duarte - carlosduarte@imedea.uib-csic.es}%
      }


\address{%
    \iid(1)Instituto de F\'isica Interdisciplinar y Sistemas Complejos, IFISC (CSIC-UIB),
Campus Universitat de les Illes Balears, E-07122 Palma de Mallorca, Spain\\
    \iid(2)Instituto Mediterraneo de Estudios Avanzados, IMEDEA (CSIC-UIB),
C/ Miquel Marqu\'es 21, E-07190 Esporles, Spain\\
    \iid(3)Oceans Institute, University of Western Australia,
35 Stirling Highway, Crawley 6009, Australia
}%

\maketitle


\begin{abstract}
        \paragraph*{Background:} One of the classical questions
in evolutionary biology is how evolutionary processes are coupled
at the gene and species level. With this motivation, we compare
the topological properties (mainly the depth scaling, as a
characterization of balance) of a large set of protein phylogenies
with those of a set of species phylogenies.

        \paragraph*{Results:} The comparative analysis between
protein and species phylogenies shows that both sets of
phylogenies share a remarkably similar scaling behavior,
suggesting the universality of branching rules and of the
evolutionary processes that drive biological diversification
from gene to species level. In order to explain such
generality, we propose a simple model which allows us to
estimate the proportion of evolvability/robustness needed to
approximate the scaling behavior observed in the phylogenies,
highlighting the relevance of the robustness of a biological
system (species or protein) in the scaling properties of the
phylogenetic trees.

        \paragraph*{Conclusions:} The invariance of the scaling
properties at levels spanning from genes to species suggests
that rules that govern the incapability of a biological
system to diversify are equally relevant both at the gene and
at the species level.
\end{abstract}

\ifthenelse{\boolean{publ}}{\begin{multicols}{2}}{}


\section*{Background}
 During the last century, an important effort has been devoted
to the understanding of diversification patterns and processes
in terms of branching evolutionary trees
\cite{willis22,savage83,burlando93,kirkpatrick93,mooers97,blum06,herrada08}.
Tempo and mode of genetic change, and their connections with
tempo and mode of speciation is an important issue in this
context. In that sense, we address the question of whether
similar forces act across the gene level and species-level
evolution \cite{morris00,carroll05,roth07}, through a
comparative analysis of the topological behavior of protein and
species phylogenies.

Previous analyses of the topological properties of phylogenies
have revealed universal patterns of phylogenetic
differentiation \cite{dial89, burlando90, burlando93, blum06,
herrada08}. This means that the impact of evolutionary forces
shaping the diversity of life on Earth on the shape of
phylogenetic trees is, at least to the level of detail captured
by the descriptors used, similar across a broad range of
scales, from macro-evolution to speciation and population
differentiation, and across diverse organisms such as
eukaryotes, eubacteria, archaea or viruses, thereby. This
together with the fact that evolutionary forces work at
molecular level motivates the study of the topology of
evolutionary relationships among molecular entities, looking
for patterns of differentiation at such molecular level,
thereby extending the examination of the universality of the
scaling of branching laws in phylogenies all the way from
molecular- to macro-evolution.

The term ``protein family" was coined by Dayhoff in the 1960's
to comprise similar proteins in structure and/or function,
which are presumed to have evolved from a common ancestor
protein \cite{dayhoff65}. Our analysis is based on a thorough
data set of 7,738 protein families downloaded from the PANDIT
database ({\tt http://www.ebi.ac.uk/goldman-srv/pandit/})
\cite{whelan06} on May 27th 2008. It contains families with
a broad range of sizes (see Figure 1). Taking into account
that protein family diversification is driven by alternative
evolutionary processes beyond speciation (orthology), such as
gene duplication (paralogy), these data were used to test if
the universal patterns found previously in species, subspecies,
and higher taxonomic levels, also apply at the molecular
evolutionary level. In particular we use tools derived from
modern network theory
\cite{banavar99,garlaschelli03,camacho05,klemm05,herrada08} to
examine the scaling of the branching in the protein family
phylogenies.

A protein family phylogeny is represented as a tree, \ie, as an
acyclic graph of nodes connected by branches (links), where
each node represents a diversification event. For each node in
a phylogeny, a subtree (or subfamily) $S$ is made of the root
at the selected node and all of its descendant nodes. The
subtree size $A$ is the total number of subfamily members that
diversify from the root (including itself). The
characterization of how protein diversity is arranged through
the phylogenies can be achieved in a variety of ways
\cite{apic03,unger03,cotton05,kunin05,lee05,cotton06,sales07,hughes08}.
We focus here on the {\em mean depth}, $d$, of the subtree $S$
(see Methods) \cite{ford06,blum06,hernandez10} defined as: $d
= \sum_j d_{\textrm{\scriptsize{root}},j}/A$, where, for a
given node $j$, $d_{\textrm{\scriptsize{root}},j}$ is its
topological distance to the root of the subtree $S$, that is,
the number of nodes one has to go through so as to go from that
node to the root (including the root in the counting), and the
sum is over all nodes in the subtree $S$. Note that we use here
the mean depth over all subtree nodes and not just the leaves,
which gives a different but related measure \cite{sackin72,
kirkpatrick93, blum05}. In the remainder, when no subindex is
indicated, we understand that mean depth and other quantities
refer to a whole tree or a subtree depending on the context.

How the shape of a phylogenetic tree, \ie, the distribution of
protein diversification, changes with tree size, \ie, with the
number of proteins it contains, can be analyzed by examining
the dependence of the mean depth on subfamily size $d=d(A)$.
This gives information on the balance characteristics of the
tree. To be clearer, in the additional file 1 we show the
analysis of $A$ and $d$ for a fully balanced and a fully imbalanced
15-tip tree, as well as for a 15-tip subtree of a real phylogenetic tree.
For a given tree size, the smallest value of the mean depth
corresponds to the fully polytomic tree. The mean depth $d$ as
a function of tree size $A$ is given in this case by
\be
 d_\textrm{\scriptsize{min}} = 1- \frac{1}{A}~.
\ee
For large sizes the leading contribution is
$d_\textrm{\scriptsize{min}} \sim 1$. The largest mean depth
value for a given size is given by the fully imbalanced, or
asymmetric, binary tree with a mean depth given by
\be
d_\textrm{\scriptsize{max}} = \frac{1}{4}\bigg(\frac{A^2-1}{A}\bigg)~,
\ee
which for large sizes $A$ leads to the scaling behavior
$d_\textrm{\scriptsize{max}} \sim A$. The fully balanced, or
symmetric, binary tree is inside these extremes, with a mean
depth given by
\be
d = \frac{((A+1)\ln_2(A+1)-2A)}{A}~.
\ee
The leading contribution at large sizes is logarithmic: $d \sim
\ln A$. This logarithmic scaling is not exclusive of fully
balanced trees, it is also the behavior of the Equal-Rates
Markov (ERM) model \cite{cavalli67,harding71,hernandez10}, the
natural null model for stochastic tree construction, in which,
at each time step, one of the existing leaves of the tree is
chosen at random and bifurcated into two new leaves.

We report here the patterns of mean depth for protein families,
and compare the branching patterns derived for protein
families, from the PANDIT database with those of species
phylogenies, reported previously from the TreeBASE database
\cite{herrada08}. This comparison shows that branching patterns
are mostly preserved across evolutionary scales spanning from
genes to species.


%
%
%

\section*{Results}
  \subsection*{Protein phylogenies depth scaling}
The analysis of the 7,738 protein phylogenies of PANDIT
database shows (Figure 2) that the scaling of the mean depth
with tree size lies between the two extreme topologies for
binary trees (fully imbalanced and fully balanced trees),
with the exception of a few polytomic subtrees, which display
mean depth values lower that the one expected for the same size
fully balanced binary tree. The data for independent protein
trees are not scattered between the extreme cases but instead
cluster in a space intermediate between these extremes
depending on the size of the trees. Figure 2 displays
depth, averaged within logarithmic bins of values of tree size
$A$, as a function of $A$. The axes of this and other plots in
the following have been chosen so that a depth behavior of the
form $d \sim (\ln A)^2$ will appear as a straight line. This is
the behavior suggested by the models in
\cite{aldous01,keller11} and for organisms phylogenies
in \cite{blum06}, which seems to correspond rather well
to our data. The fully imbalanced tree shows a linear
dependence $d \sim A$, and the fully balanced tree shows a
logarithmic dependence of the form $d \sim \ln A$ (lines also
shown in Figure 2). 

We analyzed the scaling of the mean depth as function of the
tree size for different protein functions (e.g. nuclear,
structural, metabolic) to assess whether different protein
functions show scaling laws departing from the average mean
depth scaling described for the whole PANDIT database. The
results obtained show that the depth of different protein
functions shows the same scaling with tree size as that
described for the whole PANDIT dataset independently of
function (Figure 3). This result supports the existence of
universal scaling laws in the depth of protein phylogenies.

The universality observed in the depth scaling of protein
phylogenies is even more remarkable when protein phylogenies
are compared with the species phylogenies \cite{herrada08}
obtained from the TreeBASE database (Figure 4). The comparative
analysis between PANDIT and TreeBASE shows a similar scaling of
the mean depth with the tree size for both datasets. Although
in a previous work with organism phylogenies the depth scaling
was fitted to a power law \cite{herrada08}, we find here
that the squared logarithmic scaling $d \sim (\ln A)^2$ of
\cite{aldous01,blum06,keller11} provides also a
reasonable fit for the protein families. Discriminating between
these two scaling laws requires the comparison of larger trees,
which are not available at the moment. Further discussion on
this point is provided in the additional file 2. The important
point, however, is that the analysis of protein phylogenies
shows that the trees follow a scaling law as they speciate,
which is universal across protein functions, and similar to
that associated with the speciation at the species level.

There is some dispersion of the mean depth for the whole PANDIT
dataset observed in Figure 2, which is attributable to
imbalanced bifurcations in some specific trees. This increase
in the presence of imbalanced bifurcations is reflected as a
fast increase, characteristic of fully imbalanced trees. These
regions with a high number of imbalanced bifurcations are most
of the times close to the root, which can be related to a lack
of resolution in the reconstruction process. In Figure 5 we
show a detailed example of a phylogenetic tree with a region
with a high presence of imbalance in the bifurcations close to
the root, that leads to a dispersion from the mean depth
scaling in the range $A\in \left(2\times10^2,
3\times10^2\right)$, preserving the previously described
universal mean depth scaling behavior in most of the size
range, from $1$ to $2\times10^2$. The fact that the deviation
from the mean is restricted only to certain regions of the
phylogenetic trees, and that they do not affect
significatively the average depth, thus preserving the global
trend, supports the overall universality of the average depth
scaling behavior found in the protein phylogenies from the
PANDIT database.

  \subsection*{Evolvability model}
The depth scaling behavior shared by protein and species
phylogenies can be explained by different branching mechanisms.
In this direction, during the last decade, several models have
been published proposing different mechanisms to capture the
topology of phylogenetic trees
\cite{aldous01,pinelis03,blum06,ford06,stich09,hernandez10}.
Most of the models proposed yield a logarithmic scaling of the
mean depth, \ie, ERM-type for large sizes
\cite{yule24,cavalli67,harding71}, which is not a good
description of our data (see Figure 2 and additional file 2), at least at the tree sizes
available; the AB model proposed in Ref.~\cite{aldous01} is one of
the few models that deviate from the ERM-like scaling leading
to a squared logarithmic $d \sim (\ln A)^2$ (see also
\cite{blum06}); models with power law scaling of the mean depth
$d\sim A^\eta$ have also been defined in terms of
statistical rules assigning probabilities to different splittings or types of
trees \cite{aldous01} or in terms of (simplified)
evolutionary events (in the sense specified in Ref.~\cite{pinelis03}) occurring in time
\mbox{\cite{ford06,hernandez10}}.

An alternative explanation of the scaling properties of the
phylogenetic trees \cite{stich09} suggests that the non-ERM
behavior is a small-size transient behavior, which would
cross-over to the ERM scaling $d \sim \ln A$ as larger tree
sizes become available.

The process conducive to trees that deviate from ERM behavior
is the presence of temporal correlations, which leads to
asymptotic or just finite-size deviations with respect to the
ERM behavior depending on whether these correlations are
permanent or restricted to finite but large times. We, thus,
explored the role of such correlations through a simple model
based on the inheritability of the evolvability, \ie, the
ability to evolve \cite{dawkins89,brookfield09}, as a
biological characteristic which is itself inherited by sister
species in speciation events. The process starts with the root,
which we consider capable to speciate. At each time step,
all present species capable to speciate branch simultaneously.
Each branching event yields two new daughter species, for
which we allow two possible outcomes:

\begin{itemize}
 \item  with probability $p$, the new species inherit the evolvability
 of the mother species, \ie, they have the same capacity as the mother species to speciate again;
 \item with probability $1-p$, one of the daughter species is unable to speciate
 again, that is, only one of the two daughter species preserves the ability to evolve.
 Stemming from the definition of \textit{robustness} as the property of a system to
 remain invariant in the presence of genetic or environmental perturbations \cite{masel09},
 we consider a species' inability to speciate its robustness.
\end{itemize}

The first case gives rise to a symmetric speciation event, in
which the two species emerging from the speciation event are
similar, while the second one giving rise to asymmetries in the
tree. If $p = 1$, we recover the completely balanced binary
tree, while the topology obtained in the other extreme, $p =
0$, is the completely imbalanced binary tree
(Figure 6). Thus the model combines symmetric
with asymmetric branching introducing correlations (since one
occurrence of the asymmetric event precludes further speciation
on that branch), with the proportion determined by the
parameter $p$.

The trees generated with this algorithm yield a scaling very
close to those observed for phylogenetic trees in both PANDIT
and TreeBASE for $p=0.24$ (see Figure 6, and additional
file 3). This result identifies the prevalence of imbalanced
branching events (occurring with frequency $1-p=0.76$) relative
to balanced ones ($p=0.24$), which is consistent with earlier
reports \cite{mooers97,aldous01,blum06}.

The correlations introduced by our model are not, however,
permanent and ultimately a crossover to the random behavior
appears for long sizes. To evaluate this, we calculated the
analytical expression of the average depth, $d$. Taking into
account that the expected number of offsprings of a pair of
sister nodes is $2z=4p+2(1-p)=2(1+p)$, starting with the root,
the expected number of nodes after $n$ branching events is
\be
A = 1+2 \bigg[\sum_{i=0}^{n-1} z^i\bigg]
= 1+2\frac{z^n-1}{z-1}~,
\ee
where  $z=1+p$ is the expected offspring per sister node. The
expected value of the cumulative branch size (see Methods) is given by
\ba
C &=&
1+2 \bigg[\sum_{i=0}^{n-1} z^i(i+2)\bigg] \\ \nonumber
&=&  1+2\bigg\{z \frac{(n-1)z^n-nz^{n-1}+1}{(z-1)^2}+ 2\frac{z^n-1}{z-1}\bigg\}~.
\ea
At large $n$, the leading contributions are $A\sim z^n$ and
$C\sim n z ^n$ (we do not write explicitly prefactors which may
depend on $z$ but not on $n$). Taking into account Eq.~(7) in
Methods (i.e. $d=(C/A)-1$) and inverting the relationship
between $A$ and $n$ ($n \sim \ln A$), we obtain that for large
sizes the leading order of the mean depth is $d \sim \ln A$,
which indicates that what we observe in the simulations is a
long transient behavior. This transient behavior leads to the
fact that our model fits the proper behavior of the data at
the sizes in the databases, but the asymptotic scaling at the
larger sizes will finally be $d \sim \ln A$, as in the ERM.


\section*{Discussion}
The development of high-throughput ``-omics'' has provided the
data required to address the traditional debate on how
gene-level evolution shapes the species-level evolution
\cite{morris00,carroll05,roth07}. This debate connects with
that on the (dis)continuity between micro- and macro-evolution,
and gradualism versus saltationism
\cite{erwin00,simons02,grantham07}. In the context of these
debates, the universal scaling of phylogenetic trees at intra-
and inter-specific levels shown earlier \cite{herrada08}
suggested the conservation of the evolutionary processes that
drive biological diversification across the entire history of
life. Here we extend this observation further to demonstrate
that the universality of the scaling properties can also be
extrapolated to the gene-level. The results presented here show
that the branching and scaling patterns in protein families do
not differ significantly from the patterns observed in species
phylogenies, at least for the topological properties we have
calculated. We do not observe any discrepancy between the shape
of protein phylogenies and species phylogenies. Moreover, the
results presented here shows no evidence for possible
differences in phylogenetic trees among protein families with
different biological functions, further providing evidence of
universal, conserved evolutionary processes from genes to
species.

In 2006, Cotton and Page published a comparative analysis
between human gene phylogenies and species phylogenies
\cite{cotton06}. They found quantitative differences between
human paralogous gene and orthologous gene phylogenies. Their
research focused on the comparison between (small) paralogous
and orthologous gene families, while here we have analyzed
complete protein families, which included both paralogous and
orthologous protein members, focusing on the comparison between
protein and species phylogenies. Our approach is based
on a scaling analysis, examining how variables change with tree
size, whereas the Cotton-Page's approach is based on a
quantitative analysis of small sizes. This implies that despite
their finding of quantitative differences between paralogous
and orthologous gene phylogenies, we expect that both
phylogenies would display scaling behavior similar to that we
described here for complete protein phylogenies and organism
phylogenies \cite{herrada08}.

Different evolutionary models and mechanisms have been proposed
to explain the branching patterns arising in evolution
\cite{yule24,aldous01,blum06,ford06,stich09,hernandez10}. Here
we have introduced a simple model accounting for differences in
the degree of \textit{evolvability}, which is emerging as a key
trait constraint as important as robustness in evolution
\cite{wagner05, lenski06, daniels08,wagner08}. The model we
proposed can be interpreted in the framework of the balance
between evolvability as the potential of a biological system
for future adaptive mutation and evolution
\cite{brookfield09}, and robustness as the property of a
system to produce relatively invariant output in the presence
of a perturbation \cite{masel09}. Indeed, the symmetric
diversification event should correspond to the biological
context in which the biological system is evolvable, while the
asymmetric diversification process should correspond to a
biological context where the new biological system, which has
just appeared from the diversification process, is robust and
unable of unlimited diversification.

The asymptotic behavior of our model at long tree sizes
recovers the logarithmic behavior of the ERM scaling, so that,
as in the models by \cite{stich09}, the non-ERM behavior occurs
as a transient for the relatively small tree sizes present
in the databases. Despite this, the local (i.e. present for
finite sizes) imbalance in real trees can be interpreted in
terms of the \textit{evolvability} concept. The prevalence of
rhe unbalanced branching found is consistent with previous
works \cite{guyer91, heard92, guyer93, mooers95, aldous01,
blum06}, and has been traditionally explained by the presence
of variations in the speciation and/or extinction rates
throughout the Tree of Life \cite{kirkpatrick93, mooers97}.
Different biological explanations for these variations in the
speciation and/or extinction rates have been proposed, such as:
refractory period \cite{chan99}, mass extinctions
\cite{heard02}, specialization \cite{kirkpatrick93} or
environment effects \cite{davies05}. The consideration of an
evolutionary scenario based on the evolvability/robustness
interplay has led us to postulate the presence of asymmetric
diversification events over the depth scaling during
evolutionary processes giving rise to a new biological system
which is unable to undergo a new diversification event. An
incapability to diversify may occur at different levels of
evolution, and can be found at the macroevolutionary level with
taxa that require very long refractory periods or with random
massive extinctions of taxa, as well as at the
microevolutionary or gene level, where the elements unable to
diversify are individuals from a population or genetic variants
from a cell, embryo or individual.

\section*{Conclusions}
In summary, the finding of universal scaling properties at gene
and species level, characterized by the similar scaling laws,
strongly suggest the universality of branching rules, and hence
of the evolutionary processes that drive biological
diversification across the entire history of life, from genes
to species. The topological characterization of phylogenetic
trees has proven helpful to analyse the relevance of the
robustness of a biological system (species or protein) in the
scaling properties of the phylogenetic trees. Thus, the
invariance of the scaling properties at levels spanning from
genes to species suggests that the mechanisms leading to the
incapability of a biological system to diversify for a very
long period of time act at both the gene- and species-level.

\section*{Methods}
\subsection{Protein phylogenies database}
We analyzed the 7,738 protein families available in the PANDIT
database (\url{http://www.ebi.ac.uk/goldman-srv/pandit/}
accession date May 27, 2008) \cite{whelan06}. PANDIT is based
upon Pfam ({\tt http://pfam.sanger.ac.uk/}) \cite{bateman04},
and constitutes a large collection of protein family
phylogenies from different signalling pathways, cellular
organelles and biological functions, reconstructed with
five different methods: NJ\cite{saitou87},
BioNJ\cite{gascuel97}, Weighbor\cite{bruno00},
FastME\cite{desper02} and Phyml\cite{guindon03}. The size
of each of the protein phylogenies, $T$, ranges from 2 to
more than 2000 tips (i.e. proteins within families) and, in
agreement with previous reports \cite{huynen98, harrison02,
koonin02, luscombe02, unger03}, shows a power law distribution
$P(T)\sim T^{-\gamma}$ (see Figure 1).  Most of the
bifurcations in these phylogenies are binary, with only 22\% of
polytomic bifurcations.

\subsection{Mean depth}
The definition of the mean depth $d$ used here is directly
related to the cumulative branch size
\cite{garlaschelli03,campos04,camacho05,klemm05,herrada08}
defined as $C = \sum_j A_j$. The sum runs over all nodes $j$ in
a tree and $A_j$ corresponds to the size of the subtree $S_j$.
The relationship between $C$ and the mean depth can be obtained
taking into account that the cumulative branch size can also be
written as
\be
C= \sum_j {(d_{\textrm{\scriptsize{root}},j}+1)} =d A +  A~,
\ee
where $d_{\textrm{\scriptsize{root}},j}$ is the distance of
node $j$ to the root. Thus, the mean depth of a tree is
obtained as
\begin{equation}
d = \frac{C}{A} - 1~.
\label{CdA}
\end{equation}

The depth of a tree can also be characterized by taking into
account only the distance from the tips to the root. This is
the case of the Sackin's index, $S$, which is defined as the
sum of the depths of all the leaves of the tree $S = \sum_j
d_{\textrm{\scriptsize{root}},j}$ \cite{sackin72}. Taking into
account that a binary tree can be obtained as a growing tree
adding at each time a speciation event we can calculate the
change $\Delta C$ and $\Delta S$ at each speciation. If the
distance of the node that speciates (leading to two new nodes)
to the root is $d'$ then
\be
 \Delta C = 2(d' +2) = 2d' +4 ~,
\ee
while
\be
\Delta S= -d' + 2(d'+1) = d'+2~.
\ee
Accounting for the initial condition, that is, the root, with
$C=1$ and $S=0$, yields $C=2S+1$ for binary trees. Thus, at
large sizes, both quantities, $C$ and $S$, become proportional
and scale in the same way with size.


\section*{Authors contributions}
EAH downloaded the protein phylogenies database, carried out the
depth scaling analysis and designed the evolvability model.
EAH, VME and EHG designed the depth scaling analysis as well as
provided the mathematical framework of the work. VME, EHG and CMD
supervised the work. All authors participated in planning
the work and writing the manuscript and read and
approved the final manuscript.

\section*{Acknowledgements}
  \ifthenelse{\boolean{publ}}{\small}{}
We acknowledge financial support from the European Commission
through the NEST-Complexity project EDEN (043251) and from MICINN
(Spain) and FEDER through project FISICOS (FIS2007-60327).



\newcommand{\BMCxmlcomment}[1]{}

\BMCxmlcomment{

<refgrp>

<bibl id="B1">
  <title><p>Age and area: a study in geographical distribution and origin of
  species</p></title>
  <aug>
    <au><snm>Willis</snm><fnm>J. C.</fnm></au>
  </aug>
  <publisher>Cambridge: Cambridge University Press</publisher>
  <pubdate>1922</pubdate>
</bibl>

<bibl id="B2">
  <title><p>The shape of evolution: systematic tree topology</p></title>
  <aug>
    <au><snm>Savage</snm><fnm>H. M.</fnm></au>
  </aug>
  <source>Biol. J. Linnean Soc.</source>
  <pubdate>1983</pubdate>
  <volume>20</volume>
  <fpage>225</fpage>
  <lpage>-244</lpage>
</bibl>

<bibl id="B3">
  <title><p>The fractal geometry of evolution.</p></title>
  <aug>
    <au><snm>Burlando</snm><fnm>B</fnm></au>
  </aug>
  <source>J. Theor. Biol.</source>
  <pubdate>1993</pubdate>
  <volume>163</volume>
  <issue>2</issue>
  <fpage>161</fpage>
  <lpage>-172</lpage>
</bibl>

<bibl id="B4">
  <title><p>Searching for Evolutionary Patterns in the Shape of a Phylogenetic
  Tree</p></title>
  <aug>
    <au><snm>Kirkpatrick</snm><fnm>M.</fnm></au>
    <au><snm>Slatkin</snm><fnm>M.</fnm></au>
  </aug>
  <source>Evolution</source>
  <pubdate>1993</pubdate>
  <volume>47</volume>
  <fpage>1171</fpage>
  <lpage>-1181</lpage>
</bibl>

<bibl id="B5">
  <title><p>Inferring evolutionary process from the phylogenetic tree
  shape</p></title>
  <aug>
    <au><snm>Mooers</snm><fnm>A. O.</fnm></au>
    <au><snm>Heard</snm><fnm>S. B.</fnm></au>
  </aug>
  <source>Q. Rev. Biol.</source>
  <pubdate>1997</pubdate>
  <volume>72</volume>
  <fpage>31</fpage>
  <lpage>-54</lpage>
</bibl>

<bibl id="B6">
  <title><p>Which random processes describe the tree of life? A large-scale
  study of phylogenetic tree imbalance.</p></title>
  <aug>
    <au><snm>Blum</snm><fnm>MGB</fnm></au>
    <au><snm>Fran\c{c}ois</snm><fnm>O</fnm></au>
  </aug>
  <source>Syst. Biol.</source>
  <pubdate>2006</pubdate>
  <volume>55</volume>
  <issue>4</issue>
  <fpage>685</fpage>
  <lpage>-691</lpage>
</bibl>

<bibl id="B7">
  <title><p>Universal Scaling in the Branching of the Tree of Life</p></title>
  <aug>
    <au><snm>Herrada</snm><fnm>E. A.</fnm></au>
    <au><snm>Tessone</snm><fnm>C. J.</fnm></au>
    <au><snm>Klemm</snm><fnm>K.</fnm></au>
    <au><snm>Egu{\'i}luz</snm><fnm>V. M.</fnm></au>
    <au><snm>Hern{\'a}ndez Garc{\'i}a</snm><fnm>E.</fnm></au>
    <au><snm>Duarte</snm><fnm>C. M.</fnm></au>
  </aug>
  <source>PLoS ONE</source>
  <pubdate>2008</pubdate>
  <volume>3</volume>
  <fpage>e2757</fpage>
</bibl>

<bibl id="B8">
  <title><p>Evolution: bringing molecules into the fold</p></title>
  <aug>
    <au><snm>Morris</snm><fnm>S. C.</fnm></au>
  </aug>
  <source>Cell</source>
  <pubdate>2000</pubdate>
  <volume>100</volume>
  <fpage>1</fpage>
  <lpage>-11</lpage>
</bibl>

<bibl id="B9">
  <title><p>Evolution at two levels: on genes and form.</p></title>
  <aug>
    <au><snm>Carroll</snm><fnm>SB</fnm></au>
  </aug>
  <source>PLoS Biol.</source>
  <pubdate>2005</pubdate>
  <volume>3</volume>
  <issue>7</issue>
  <fpage>e245</fpage>
</bibl>

<bibl id="B10">
  <title><p>Evolution after gene duplication: models, mechanisms, sequences,
  systems, and organisms.</p></title>
  <aug>
    <au><snm>Roth</snm><fnm>C</fnm></au>
    <au><snm>Rastogi</snm><fnm>S</fnm></au>
    <au><snm>Arvestad</snm><fnm>L</fnm></au>
    <au><snm>Dittmar</snm><fnm>K</fnm></au>
    <au><snm>Light</snm><fnm>S</fnm></au>
    <au><snm>Ekman</snm><fnm>D</fnm></au>
    <au><snm>Liberles</snm><fnm>DA</fnm></au>
  </aug>
  <source>J. Exp. Zool. B Mol. Dev. Evol.</source>
  <pubdate>2007</pubdate>
  <volume>308</volume>
  <issue>1</issue>
  <fpage>58</fpage>
  <lpage>-73</lpage>
</bibl>

<bibl id="B11">
  <title><p>Nonrandom diversification within taxonomic assemblages</p></title>
  <aug>
    <au><snm>Dial</snm><fnm>K. P.</fnm></au>
    <au><snm>Marzluff</snm><fnm>J. M.</fnm></au>
  </aug>
  <source>Syst. Zool.</source>
  <pubdate>1989</pubdate>
  <volume>38</volume>
  <fpage>26</fpage>
  <lpage>-37</lpage>
</bibl>

<bibl id="B12">
  <title><p>The fractal dimension of taxonomic systems</p></title>
  <aug>
    <au><snm>Burlando</snm><fnm>B</fnm></au>
  </aug>
  <source>J. Theor. Biol.</source>
  <pubdate>1990</pubdate>
  <volume>146</volume>
  <fpage>99</fpage>
  <lpage>-114</lpage>
</bibl>

<bibl id="B13">
  <title><p>Atlas of Protein Sequence and Structure</p></title>
  <aug>
    <au><snm>Dayhoff</snm><fnm>M. O.</fnm></au>
  </aug>
  <publisher>Washington: National Biomedical Research Foundation</publisher>
  <pubdate>1965-1978</pubdate>
</bibl>

<bibl id="B14">
  <title><p>PANDIT: an evolution-centric database of protein and associated
  nucleotide domains with inferred trees</p></title>
  <aug>
    <au><snm>Whelan</snm><fnm>S</fnm></au>
    <au><snm>{de Bakker}</snm><fnm>PIW</fnm></au>
    <au><snm>Quevillon</snm><fnm>E</fnm></au>
    <au><snm>Rodriguez</snm><fnm>N</fnm></au>
    <au><snm>Goldman</snm><fnm>N</fnm></au>
  </aug>
  <source>Nucleic Acids Res.</source>
  <pubdate>2006</pubdate>
  <volume>34</volume>
  <issue>Database issue</issue>
  <fpage>D327</fpage>
  <lpage>-D331</lpage>
</bibl>

<bibl id="B15">
  <title><p>Size and form in efficient transportation networks</p></title>
  <aug>
    <au><snm>Banavar</snm><fnm>J. R.</fnm></au>
    <au><snm>Maritan</snm><fnm>A.</fnm></au>
    <au><snm>Rinaldo</snm><fnm>A.</fnm></au>
  </aug>
  <source>Nature</source>
  <pubdate>1982</pubdate>
  <volume>399</volume>
  <fpage>130</fpage>
  <lpage>-132</lpage>
</bibl>

<bibl id="B16">
  <title><p>Universal scaling relations in food webs.</p></title>
  <aug>
    <au><snm>Garlaschelli</snm><fnm>D</fnm></au>
    <au><snm>Caldarelli</snm><fnm>G</fnm></au>
    <au><snm>Pietronero</snm><fnm>L</fnm></au>
  </aug>
  <source>Nature</source>
  <pubdate>2003</pubdate>
  <volume>423</volume>
  <issue>6936</issue>
  <fpage>165</fpage>
  <lpage>-168</lpage>
</bibl>

<bibl id="B17">
  <title><p>Food-web topology: universal scaling in food-web
  structure?</p></title>
  <aug>
    <au><snm>Camacho</snm><fnm>J</fnm></au>
    <au><snm>Arenas</snm><fnm>A</fnm></au>
  </aug>
  <source>Nature</source>
  <pubdate>2005</pubdate>
  <volume>435</volume>
  <issue>7044</issue>
  <fpage>E3</fpage>
  <lpage>-E4</lpage>
</bibl>

<bibl id="B18">
  <title><p>Scaling in the structure of directory trees in a computer
  cluster.</p></title>
  <aug>
    <au><snm>Klemm</snm><fnm>K</fnm></au>
    <au><snm>Egu{\'i}luz</snm><fnm>VM</fnm></au>
    <au><snm>{San Miguel}</snm><fnm>M</fnm></au>
  </aug>
  <source>Phys. Rev. Lett.</source>
  <pubdate>2005</pubdate>
  <volume>95</volume>
  <issue>12</issue>
  <fpage>128701</fpage>
</bibl>

<bibl id="B19">
  <title><p>Multi-domain protein families and domain pairs: comparison with
  known structures and a random model of domain recombination.</p></title>
  <aug>
    <au><snm>Apic</snm><fnm>G</fnm></au>
    <au><snm>Huber</snm><fnm>W</fnm></au>
    <au><snm>Teichmann</snm><fnm>SA</fnm></au>
  </aug>
  <source>J. Struct. Funct. Genomics</source>
  <pubdate>2003</pubdate>
  <volume>4</volume>
  <issue>2-3</issue>
  <fpage>67</fpage>
  <lpage>-78</lpage>
</bibl>

<bibl id="B20">
  <title><p>Scaling law in sizes of protein sequence families: from
  super-families to orphan genes.</p></title>
  <aug>
    <au><snm>Unger</snm><fnm>R</fnm></au>
    <au><snm>Uliel</snm><fnm>S</fnm></au>
    <au><snm>Havlin</snm><fnm>S</fnm></au>
  </aug>
  <source>Proteins</source>
  <pubdate>2003</pubdate>
  <volume>51</volume>
  <issue>4</issue>
  <fpage>569</fpage>
  <lpage>-576</lpage>
</bibl>

<bibl id="B21">
  <title><p>Rates and patterns of gene duplication and loss in the human
  genome.</p></title>
  <aug>
    <au><snm>Cotton</snm><fnm>JA</fnm></au>
    <au><snm>Page</snm><fnm>RDM</fnm></au>
  </aug>
  <source>Proc. R. Soc. B</source>
  <pubdate>2005</pubdate>
  <volume>272</volume>
  <issue>1560</issue>
  <fpage>277</fpage>
  <lpage>-283</lpage>
</bibl>

<bibl id="B22">
  <title><p>The properties of protein family space depend on experimental
  design.</p></title>
  <aug>
    <au><snm>Kunin</snm><fnm>V</fnm></au>
    <au><snm>Teichmann</snm><fnm>SA</fnm></au>
    <au><snm>Huynen</snm><fnm>MA</fnm></au>
    <au><snm>Ouzounis</snm><fnm>CA</fnm></au>
  </aug>
  <source>Bioinformatics</source>
  <pubdate>2005</pubdate>
  <volume>21</volume>
  <issue>11</issue>
  <fpage>2618</fpage>
  <lpage>-2622</lpage>
</bibl>

<bibl id="B23">
  <title><p>Identification and distribution of protein families in 120
  completed genomes using Gene3D.</p></title>
  <aug>
    <au><snm>Lee</snm><fnm>D</fnm></au>
    <au><snm>Grant</snm><fnm>A</fnm></au>
    <au><snm>Marsden</snm><fnm>RL</fnm></au>
    <au><snm>Orengo</snm><fnm>C</fnm></au>
  </aug>
  <source>Proteins</source>
  <pubdate>2005</pubdate>
  <volume>59</volume>
  <issue>3</issue>
  <fpage>603</fpage>
  <lpage>-615</lpage>
</bibl>

<bibl id="B24">
  <title><p>The shape of human gene family phylogenies.</p></title>
  <aug>
    <au><snm>Cotton</snm><fnm>JA</fnm></au>
    <au><snm>Page</snm><fnm>RDM</fnm></au>
  </aug>
  <source>BMC Evol. Biol.</source>
  <pubdate>2006</pubdate>
  <volume>6</volume>
  <fpage>66</fpage>
</bibl>

<bibl id="B25">
  <title><p>Evolution of protein families: is it possible to distinguish
  between domains of life?</p></title>
  <aug>
    <au><snm>Sales Pardo</snm><fnm>M</fnm></au>
    <au><snm>Chan</snm><fnm>AOB</fnm></au>
    <au><snm>Amaral</snm><fnm>LAN</fnm></au>
    <au><snm>Guimer{\`a}</snm><fnm>R</fnm></au>
  </aug>
  <source>Gene</source>
  <pubdate>2007</pubdate>
  <volume>402</volume>
  <issue>1-2</issue>
  <fpage>81</fpage>
  <lpage>-93</lpage>
</bibl>

<bibl id="B26">
  <title><p>The power-law distribution of gene family size is driven by the
  pseudogenisation rate's heterogeneity between gene families.</p></title>
  <aug>
    <au><snm>Hughes</snm><fnm>T</fnm></au>
    <au><snm>Liberles</snm><fnm>DA</fnm></au>
  </aug>
  <source>Gene</source>
  <pubdate>2008</pubdate>
  <volume>414</volume>
  <issue>1-2</issue>
  <fpage>85</fpage>
  <lpage>-94</lpage>
</bibl>

<bibl id="B27">
  <title><p>Probabilities on cladograms:introduction to the alpha
  model</p></title>
  <aug>
    <au><snm>Ford</snm><fnm>D. J.</fnm></au>
  </aug>
  <source>PhD thesis</source>
  <publisher>Stanford University</publisher>
  <pubdate>2006</pubdate>
</bibl>

<bibl id="B28">
  <title><p>Simple models for scaling in phylogenetic trees</p></title>
  <aug>
    <au><snm>Hern{\'a}ndez Garc{\'i}a</snm><fnm>E.</fnm></au>
    <au><snm>Tu\u{g}rul</snm><fnm>M.</fnm></au>
    <au><snm>Herrada</snm><fnm>E. A.</fnm></au>
    <au><snm>Egu{\'i}luz</snm><fnm>V. M.</fnm></au>
    <au><snm>Klemm</snm><fnm>K.</fnm></au>
  </aug>
  <source>Int. J. Bifurcat. Chaos</source>
  <pubdate>2010</pubdate>
  <volume>20</volume>
  <fpage>805</fpage>
  <lpage>-811</lpage>
</bibl>

<bibl id="B29">
  <title><p>Good and bad phenograms</p></title>
  <aug>
    <au><snm>Sackin</snm><fnm>M.</fnm></au>
  </aug>
  <source>Syst. Zool.</source>
  <pubdate>1972</pubdate>
  <volume>21</volume>
  <fpage>225</fpage>
  <lpage>-226</lpage>
</bibl>

<bibl id="B30">
  <title><p>On statistical tests of phylogenetic tree imbalance: the Sackin and
  other indices revisited.</p></title>
  <aug>
    <au><snm>Blum</snm><fnm>MGB</fnm></au>
    <au><snm>Fran\c{c}ois</snm><fnm>O</fnm></au>
  </aug>
  <source>Math. Biosci.</source>
  <pubdate>2005</pubdate>
  <volume>195</volume>
  <issue>2</issue>
  <fpage>141</fpage>
  <lpage>-153</lpage>
</bibl>

<bibl id="B31">
  <title><p>Phylogenetic analysis: models and estimation procedures</p></title>
  <aug>
    <au><snm>Cavalli Sforza</snm><fnm>L. L.</fnm></au>
    <au><snm>Edwards</snm><fnm>A. W. F.</fnm></au>
  </aug>
  <source>Am. J. Hum. Genet.</source>
  <pubdate>1967</pubdate>
  <volume>19</volume>
  <fpage>233</fpage>
  <lpage>-257</lpage>
</bibl>

<bibl id="B32">
  <title><p>The probabilities of rooted tree-shapes generated by random
  bifurcation</p></title>
  <aug>
    <au><snm>Harding</snm><fnm>E. F.</fnm></au>
  </aug>
  <source>Adv. Appl. Prob.</source>
  <pubdate>1971</pubdate>
  <volume>3</volume>
  <fpage>44</fpage>
  <lpage>-77</lpage>
</bibl>

<bibl id="B33">
  <title><p>Stochastic models and descriptive statistics for phylogenetic trees
  from Yule to today</p></title>
  <aug>
    <au><snm>Aldous</snm><fnm>D. J.</fnm></au>
  </aug>
  <source>Stat. Sci.</source>
  <pubdate>2001</pubdate>
  <volume>16</volume>
  <fpage>23</fpage>
  <lpage>-34</lpage>
</bibl>

<bibl id="B34">
  <title><p>An Age Dependent Branching Model for Macroevolution</p></title>
  <aug>
    <au><snm>Keller Schmidt</snm><fnm>S.</fnm></au>
    <au><snm>Tu\u{g}rul</snm><fnm>M.</fnm></au>
    <au><snm>Egu{\'i}luz</snm><fnm>V. M.</fnm></au>
    <au><snm>Hern{\'a}ndez Garc{\'i}a</snm><fnm>E.</fnm></au>
    <au><snm>Klemm</snm><fnm>K.</fnm></au>
  </aug>
  <source>[http://arxiv.org/abs/1012.3298]</source>
  <pubdate>2010</pubdate>
</bibl>

<bibl id="B35">
  <title><p>Evolutionary models of phylogenetic trees.</p></title>
  <aug>
    <au><snm>Pinelis</snm><fnm>I</fnm></au>
  </aug>
  <source>Proc. R. Soc. B</source>
  <pubdate>2003</pubdate>
  <volume>270</volume>
  <issue>1522</issue>
  <fpage>1425</fpage>
  <lpage>-1431</lpage>
</bibl>

<bibl id="B36">
  <title><p>Topological properties of phylogenetic trees in evolutionary
  models</p></title>
  <aug>
    <au><snm>Stich</snm><fnm>M.</fnm></au>
    <au><snm>Manrubia</snm><fnm>S. C.</fnm></au>
  </aug>
  <source>Eur. Phys. J. B</source>
  <pubdate>2009</pubdate>
  <volume>71</volume>
  <fpage>583</fpage>
  <lpage>-592</lpage>
</bibl>

<bibl id="B37">
  <title><p>A mathematical theory of evolution, based on the conclusions of Dr.
  J. C. Willis</p></title>
  <aug>
    <au><snm>Yule</snm><fnm>G. U.</fnm></au>
  </aug>
  <source>Philos. Trans. R. Soc. Lond. A</source>
  <pubdate>1924</pubdate>
  <volume>213</volume>
  <fpage>21</fpage>
  <lpage>-87</lpage>
</bibl>

<bibl id="B38">
  <title><p>The evolution of evolvability</p></title>
  <aug>
    <au><snm>Dawkins</snm><fnm>R.</fnm></au>
  </aug>
  <source>Artificial Life. The Proceedings of an Interdisciplinary Workshop on
  the Synthesis and Simulation of Living Systems, Vol. VI, September
  1987.</source>
  <publisher>Los Alamos: Addison-Wesley Pub. Corp.</publisher>
  <editor>C. Langton</editor>
  <pubdate>1989</pubdate>
  <fpage>201</fpage>
  <lpage>-220</lpage>
</bibl>

<bibl id="B39">
  <title><p>Evolution and evolvability: celebrating Darwin 200.</p></title>
  <aug>
    <au><snm>Brookfield</snm><fnm>JFY</fnm></au>
  </aug>
  <source>Biol. Lett.</source>
  <pubdate>2009</pubdate>
  <volume>5</volume>
  <issue>1</issue>
  <fpage>44</fpage>
  <lpage>-46</lpage>
</bibl>

<bibl id="B40">
  <title><p>Robustness: mechanisms and consequences.</p></title>
  <aug>
    <au><snm>Masel</snm><fnm>J</fnm></au>
    <au><snm>Siegal</snm><fnm>ML</fnm></au>
  </aug>
  <source>Trends Genet.</source>
  <pubdate>2009</pubdate>
  <volume>25</volume>
  <issue>9</issue>
  <fpage>395</fpage>
  <lpage>-403</lpage>
</bibl>

<bibl id="B41">
  <title><p>Macroevolution is more than repeated rounds of
  microevolution</p></title>
  <aug>
    <au><snm>Erwin</snm><fnm>D. H.</fnm></au>
  </aug>
  <source>Evol. Dev.</source>
  <pubdate>2000</pubdate>
  <volume>2</volume>
  <fpage>78</fpage>
  <lpage>-84</lpage>
</bibl>

<bibl id="B42">
  <title><p>The continuity of microevolution and macroevolution</p></title>
  <aug>
    <au><snm>Simons</snm><fnm>A. M.</fnm></au>
  </aug>
  <source>J. Evol. Biol.</source>
  <pubdate>2002</pubdate>
  <volume>15</volume>
  <fpage>688</fpage>
  <lpage>-701</lpage>
</bibl>

<bibl id="B43">
  <title><p>Is macroevolution more than succesive rounds of
  microevolution?</p></title>
  <aug>
    <au><snm>Grantham</snm><fnm>T.</fnm></au>
  </aug>
  <source>Paleontology</source>
  <pubdate>2007</pubdate>
  <volume>50</volume>
  <fpage>75</fpage>
  <lpage>-85</lpage>
</bibl>

<bibl id="B44">
  <title><p>Robustness and evolvability in living systems</p></title>
  <aug>
    <au><snm>Wagner</snm><fnm>A.</fnm></au>
  </aug>
  <publisher>Princeton: Princeton University Press</publisher>
  <pubdate>2005</pubdate>
</bibl>

<bibl id="B45">
  <title><p>Balancing robustness and evolvability.</p></title>
  <aug>
    <au><snm>Lenski</snm><fnm>RE</fnm></au>
    <au><snm>Barrick</snm><fnm>JE</fnm></au>
    <au><snm>Ofria</snm><fnm>C</fnm></au>
  </aug>
  <source>PLoS Biol.</source>
  <pubdate>2006</pubdate>
  <volume>4</volume>
  <issue>12</issue>
  <fpage>e428</fpage>
</bibl>

<bibl id="B46">
  <title><p>Sloppiness, robustness, and evolvability in systems
  biology.</p></title>
  <aug>
    <au><snm>Daniels</snm><fnm>BC</fnm></au>
    <au><snm>Chen</snm><fnm>YJ</fnm></au>
    <au><snm>Sethna</snm><fnm>JP</fnm></au>
    <au><snm>Gutenkunst</snm><fnm>RN</fnm></au>
    <au><snm>Myers</snm><fnm>CR</fnm></au>
  </aug>
  <source>Curr. Opin. Biotechnol.</source>
  <pubdate>2008</pubdate>
  <volume>19</volume>
  <issue>4</issue>
  <fpage>389</fpage>
  <lpage>-395</lpage>
</bibl>

<bibl id="B47">
  <title><p>Robustness and evolvability: a paradox resolved.</p></title>
  <aug>
    <au><snm>Wagner</snm><fnm>A</fnm></au>
  </aug>
  <source>Proc. R. Soc. B</source>
  <pubdate>2008</pubdate>
  <volume>275</volume>
  <issue>1630</issue>
  <fpage>91</fpage>
  <lpage>-100</lpage>
</bibl>

<bibl id="B48">
  <title><p>Comparisons between observed phylogenetic topologies with null
  expectation among three monophyletic lineages.</p></title>
  <aug>
    <au><snm>Guyer</snm><fnm>C.</fnm></au>
    <au><snm>Slowinski</snm><fnm>J. B.</fnm></au>
  </aug>
  <source>Evolution</source>
  <pubdate>1991</pubdate>
  <volume>45</volume>
  <fpage>340</fpage>
  <lpage>-350</lpage>
</bibl>

<bibl id="B49">
  <title><p>Patterns in tree balance among cladistic, phenetic, and randomly
  generated phylogenetic trees</p></title>
  <aug>
    <au><snm>Heard</snm><fnm>S. B.</fnm></au>
  </aug>
  <source>Evolution</source>
  <pubdate>1992</pubdate>
  <volume>46</volume>
  <fpage>1818</fpage>
  <lpage>-1826</lpage>
</bibl>

<bibl id="B50">
  <title><p>Adaptive radiation an the topology of large
  phylogenies.</p></title>
  <aug>
    <au><snm>Guyer</snm><fnm>C.</fnm></au>
    <au><snm>Slowinski</snm><fnm>J. B.</fnm></au>
  </aug>
  <source>Evolution</source>
  <pubdate>1993</pubdate>
  <volume>47</volume>
  <fpage>253</fpage>
  <lpage>-263</lpage>
</bibl>

<bibl id="B51">
  <title><p>Phylogenetic noise leads to unbalanced cladistic trees
  reconstructions.</p></title>
  <aug>
    <au><snm>Mooers</snm><fnm>A{\O}</fnm></au>
    <au><snm>Page</snm><fnm>RDM</fnm></au>
    <au><snm>Purvis</snm><fnm>A.</fnm></au>
    <au><snm>Harvey</snm><fnm>P. H.</fnm></au>
  </aug>
  <source>Syst. Biol.</source>
  <pubdate>1995</pubdate>
  <volume>44</volume>
  <fpage>332</fpage>
  <lpage>-342</lpage>
</bibl>

<bibl id="B52">
  <title><p>Accounting for mode of speciation increases power and realism of
  tests of phylogenetic asymmetry</p></title>
  <aug>
    <au><snm>Chan</snm><fnm>K. M. A.</fnm></au>
    <au><snm>Moore</snm><fnm>B. R.</fnm></au>
  </aug>
  <source>Am. Nat.</source>
  <pubdate>1999</pubdate>
  <volume>153</volume>
  <fpage>332</fpage>
  <lpage>-346</lpage>
</bibl>

<bibl id="B53">
  <title><p>Signatures of random and selective mass extinctions in phylogenetic
  tree balance.</p></title>
  <aug>
    <au><snm>Heard</snm><fnm>SB</fnm></au>
    <au><snm>Mooers</snm><fnm>A{\O}</fnm></au>
  </aug>
  <source>Syst. Biol.</source>
  <pubdate>2002</pubdate>
  <volume>51</volume>
  <issue>6</issue>
  <fpage>889</fpage>
  <lpage>-897</lpage>
</bibl>

<bibl id="B54">
  <title><p>Environment, area, and diversification in the species-rich
  flowering plant family Iridaceae.</p></title>
  <aug>
    <au><snm>Davies</snm><fnm>TJ</fnm></au>
    <au><snm>Savolainen</snm><fnm>V</fnm></au>
    <au><snm>Chase</snm><fnm>MW</fnm></au>
    <au><snm>Goldblatt</snm><fnm>P</fnm></au>
    <au><snm>Barraclough</snm><fnm>TG</fnm></au>
  </aug>
  <source>Am. Nat.</source>
  <pubdate>2005</pubdate>
  <volume>166</volume>
  <issue>3</issue>
  <fpage>418</fpage>
  <lpage>-425</lpage>
</bibl>

<bibl id="B55">
  <title><p>The Pfam protein families database.</p></title>
  <aug>
    <au><snm>Bateman</snm><fnm>A</fnm></au>
    <au><snm>Coin</snm><fnm>L</fnm></au>
    <au><snm>Durbin</snm><fnm>R</fnm></au>
    <au><snm>Finn</snm><fnm>RD</fnm></au>
    <au><snm>Hollich</snm><fnm>V</fnm></au>
    <au><snm>Griffiths Jones</snm><fnm>S</fnm></au>
    <au><snm>Khanna</snm><fnm>A</fnm></au>
    <au><snm>Marshall</snm><fnm>M</fnm></au>
    <au><snm>Moxon</snm><fnm>S</fnm></au>
    <au><snm>Sonnhammer</snm><fnm>ELL</fnm></au>
    <au><snm>Studholme</snm><fnm>DJ</fnm></au>
    <au><snm>Yeats</snm><fnm>C</fnm></au>
    <au><snm>Eddy</snm><fnm>SR</fnm></au>
  </aug>
  <source>Nucleic Acids Res.</source>
  <pubdate>2004</pubdate>
  <volume>32</volume>
  <issue>Database issue</issue>
  <fpage>D138</fpage>
  <lpage>-D141</lpage>
</bibl>

<bibl id="B56">
  <title><p>The neighbor-joining method: a new method for reconstructing
  phylogenetic trees.</p></title>
  <aug>
    <au><snm>Saitou</snm><fnm>N</fnm></au>
    <au><snm>Nei</snm><fnm>M</fnm></au>
  </aug>
  <source>Mol. Biol. Evol.</source>
  <pubdate>1987</pubdate>
  <volume>4</volume>
  <issue>4</issue>
  <fpage>406</fpage>
  <lpage>-425</lpage>
  <url>http://www.ncbi.nlm.nih.gov/pubmed/3447015</url>
</bibl>

<bibl id="B57">
  <title><p>BIONJ: an improved version of the NJ algorithm based on a simple
  model of sequence data.</p></title>
  <aug>
    <au><snm>Gascuel</snm><fnm>O</fnm></au>
  </aug>
  <source>Mol. Biol. Evol.</source>
  <pubdate>1997</pubdate>
  <volume>14</volume>
  <issue>7</issue>
  <fpage>685</fpage>
  <lpage>-695</lpage>
  <url>http://www.ncbi.nlm.nih.gov/pubmed/9254330</url>
</bibl>

<bibl id="B58">
  <title><p>Weighted neighbor joining: a likelihood-based approach to
  distance-based phylogeny reconstruction.</p></title>
  <aug>
    <au><snm>Bruno</snm><fnm>W J</fnm></au>
    <au><snm>Socci</snm><fnm>N D</fnm></au>
    <au><snm>Halpern</snm><fnm>A L</fnm></au>
  </aug>
  <source>Mol. Biol. Evol.</source>
  <pubdate>2000</pubdate>
  <volume>17</volume>
  <issue>1</issue>
  <fpage>189</fpage>
  <lpage>-197</lpage>
  <url>http://www.ncbi.nlm.nih.gov/pubmed/10666718</url>
</bibl>

<bibl id="B59">
  <title><p>Fast and accurate phylogeny reconstruction algorithms based on the
  minimum-evolution principle.</p></title>
  <aug>
    <au><snm>Desper</snm><fnm>R</fnm></au>
    <au><snm>Gascuel</snm><fnm>O</fnm></au>
  </aug>
  <source>J. Comput. Biol.</source>
  <pubdate>2002</pubdate>
  <volume>9</volume>
  <issue>5</issue>
  <fpage>687</fpage>
  <lpage>-705</lpage>
  <url>http://www.ncbi.nlm.nih.gov/pubmed/12487758</url>
</bibl>

<bibl id="B60">
  <title><p>A simple, fast, and accurate algorithm to estimate large
  phylogenies by maximum likelihood.</p></title>
  <aug>
    <au><snm>Guindon</snm><fnm>S.</fnm></au>
    <au><snm>Gascuel</snm><fnm>O.</fnm></au>
  </aug>
  <source>Syst. Biol.</source>
  <pubdate>2003</pubdate>
  <volume>52</volume>
  <issue>5</issue>
  <fpage>696</fpage>
  <lpage>-704</lpage>
  <url>http://www.ncbi.nlm.nih.gov/pubmed/14530136</url>
</bibl>

<bibl id="B61">
  <title><p>The frequency distribution of gene family sizes in complete
  genomes.</p></title>
  <aug>
    <au><snm>Huynen</snm><fnm>M A</fnm></au>
    <au><snm>{van Nimwegen}</snm><fnm>E</fnm></au>
  </aug>
  <source>Mol. Biol. Evol.</source>
  <pubdate>1998</pubdate>
  <volume>15</volume>
  <issue>5</issue>
  <fpage>583</fpage>
  <lpage>-589</lpage>
</bibl>

<bibl id="B62">
  <title><p>Studying genomes through the aeons: protein families, pseudogenes
  and proteome evolution.</p></title>
  <aug>
    <au><snm>Harrison</snm><fnm>PM</fnm></au>
    <au><snm>Gerstein</snm><fnm>M</fnm></au>
  </aug>
  <source>J. Mol. Biol.</source>
  <pubdate>2002</pubdate>
  <volume>318</volume>
  <issue>5</issue>
  <fpage>1155</fpage>
  <lpage>-1174</lpage>
</bibl>

<bibl id="B63">
  <title><p>The structure of the protein universe and genome
  evolution.</p></title>
  <aug>
    <au><snm>Koonin</snm><fnm>EV</fnm></au>
    <au><snm>Wolf</snm><fnm>YI</fnm></au>
    <au><snm>Karev</snm><fnm>GP</fnm></au>
  </aug>
  <source>Nature</source>
  <pubdate>2002</pubdate>
  <volume>420</volume>
  <issue>6912</issue>
  <fpage>218</fpage>
  <lpage>-223</lpage>
</bibl>

<bibl id="B64">
  <title><p>The dominance of the population by a selected few: power-law
  behaviour applies to a wide variety of genomic properties.</p></title>
  <aug>
    <au><snm>Luscombe</snm><fnm>NM</fnm></au>
    <au><snm>Qian</snm><fnm>J</fnm></au>
    <au><snm>Zhang</snm><fnm>Z</fnm></au>
    <au><snm>Johnson</snm><fnm>T</fnm></au>
    <au><snm>Gerstein</snm><fnm>M</fnm></au>
  </aug>
  <source>Genome Biol.</source>
  <pubdate>2002</pubdate>
  <volume>3</volume>
  <issue>8</issue>
  <fpage>RESEARCH0040</fpage>
</bibl>

<bibl id="B65">
  <title><p>Emergence of allometric scaling in genealogical trees</p></title>
  <aug>
    <au><snm>Campos</snm><fnm>P. R. A.</fnm></au>
    <au><snm>{de Oliveira}</snm><fnm>V. M.</fnm></au>
  </aug>
  <source>Advances in Complex Systems</source>
  <pubdate>2004</pubdate>
  <volume>7</volume>
  <fpage>39</fpage>
  <lpage>-46</lpage>
</bibl>

</refgrp>
} 


\ifthenelse{\boolean{publ}}{\end{multicols}}{}



\section*{Figures}
  \subsection*{Figure~\ref{fig1} - Protein family size distribution}
      Distribution of the size of the PANDIT protein families. Black line corresponds
      to a power law $P(T) \sim T^{-\gamma}$, with a fitted exponent $\gamma = 1.6\pm0.1$. The
      inset shows the complementary cumulative distribution $F(T)$, that is, the probability of
      finding family sizes larger than $T$.

  \subsection*{Figure~\ref{fig2} - Depth scaling of protein phylogenies}
      Mean depth scaling for all protein families in the PANDIT database (solid circles,
      where each point represents a subtree) and the corresponding averaged binned depth
      (empty circles). Where the error bars are not visible we have that the standard
      error for the mean depth
      is smaller than the symbol size. The discontinuous and continuous lines correspond to the two extreme
      binary trees: fully imbalanced and fully balanced trees, respectively. The scales of
      the axes are chosen so that a behavior of the type $d\sim (\ln A)^2$
      appears as a straight line. Note that the values below the fully balanced tree
      scaling correspond to polytomic subtrees.

  \subsection*{Figure~\ref{fig3} - Depth scaling of different protein functions}
      Binned values of the mean depth for nuclear (empty squares), structural
      (solid diamonds) and metabolic (stars) protein families. The empty circles
      represent the averaged binned depth for the whole PANDIT database.

  \subsection*{Figure~\ref{fig4} - Protein vs organism phylogenies}
      Averaged and binned mean depth for organisms in TreeBASE (solid squares) and
      for protein phylogenies in PANDIT (empty circles). Where the error bars are not visible we have that the standard
      error for the mean depth
      is smaller than the symbol size.

  \subsection*{Figure~\ref{fig5} - Mean depth behavior for a specific phylogenetic tree}
      (a) Phylogenetic tree corresponding to the Probable molybdopterin binding domain
      family (PF00994), with a high presence of imbalanced bifurcations close to the root.
      (b) Mean depth scaling of Probable molybdopterin binding domain family phylogenetic
      tree, where the empty squares correspond to the protein family. Solid circles represent
      the averaged and binned set for all the protein families of PANDIT.

  \subsection*{Figure~\ref{fig6} - Depth scaling of the evolvability model}
      The mean depth scaling of the trees generated with the evolvability model for $p=1$ and for $p=0$ reproduces the mean
      depth scaling of the fully balanced (continuous line) and imbalanced binary trees
      (discontinuous line), respectively. The trees for $p=0.24$ (empty diamonds) adjust
      the average behavior of protein (empty circles) very well. The stars correspond to trees for $p=0.5$.




\section*{Additional Files}
  \subsection*{Additional file 1 (Figure~\ref{add1}) --- Branch size and mean depth examples}
    The values of the branch size, $A$ and of the mean depth, $d$, are shown (in brackets, as ($A$,$d$))
    at each node of a fully balanced 15-tip phylogenetic tree (a), a fully imbalanced 15-tip
    phylogenetic tree (b), a 15-tip subtree of a real phylogenetic tree.

 \subsection*{Additional file 2 (Figure~\ref{add2}) --- Power-law {\sl vs.} logarithmic scaling of the depth with tree size}
  We compare the local exponents of the possible scaling laws of the depth with tree size for PANDIT.
  For sizes larger than 300 fluctuations make estimations unreliable. Filled squares: For the
  power-law scaling $d \sim A^{\eta}$ the local exponent at bin $i$ is calculated as
  $\eta_i = \Delta_i \ln d / \Delta_i \ln A$, where $\Delta_i$ indicates the  difference between
  two consecutive bins, for instance $\Delta_i \ln d = \ln d(i+1) - \ln d(i)$. Empty
  diamonds:
  For the log scaling $d \sim (\ln A)^\beta$ the local exponent at bin $i$ is calculated as
  $\beta_i = \Delta_i \ln d / \Delta_i \ln \ln A$. Constant values of the local exponents, or
  values approaching a given value as sizes increase, indicate appropriateness of the
  corresponding scaling laws to describe the data. For the power-law scaling, the exponent
  is around $\eta\simeq 0.5$ and slightly decays for larger trees. For the logarithmic scaling,
  the exponent approaches 2 as larger trees are considered, indicating $d\sim (\ln
  A)^2$. The results indicate comparable
  quality of fit for both laws at the reliable range. Note that the simpler logarithmic law, $\beta=1$, is
  not supported by the available data. 

  \subsection*{Additional file 3 (Figure~\ref{add3}) --- Standard deviation of the evolvability model}
    Values of the standard error (SE) of the results from simulations of the evolvability model
    with respect to the PANDIT dataset, for values of $p$ between $[0.21-0.27]$. A value $p=0.24$
    minimizes the error.

\begin{figure}
 \centering
\includegraphics[width=1\textwidth]{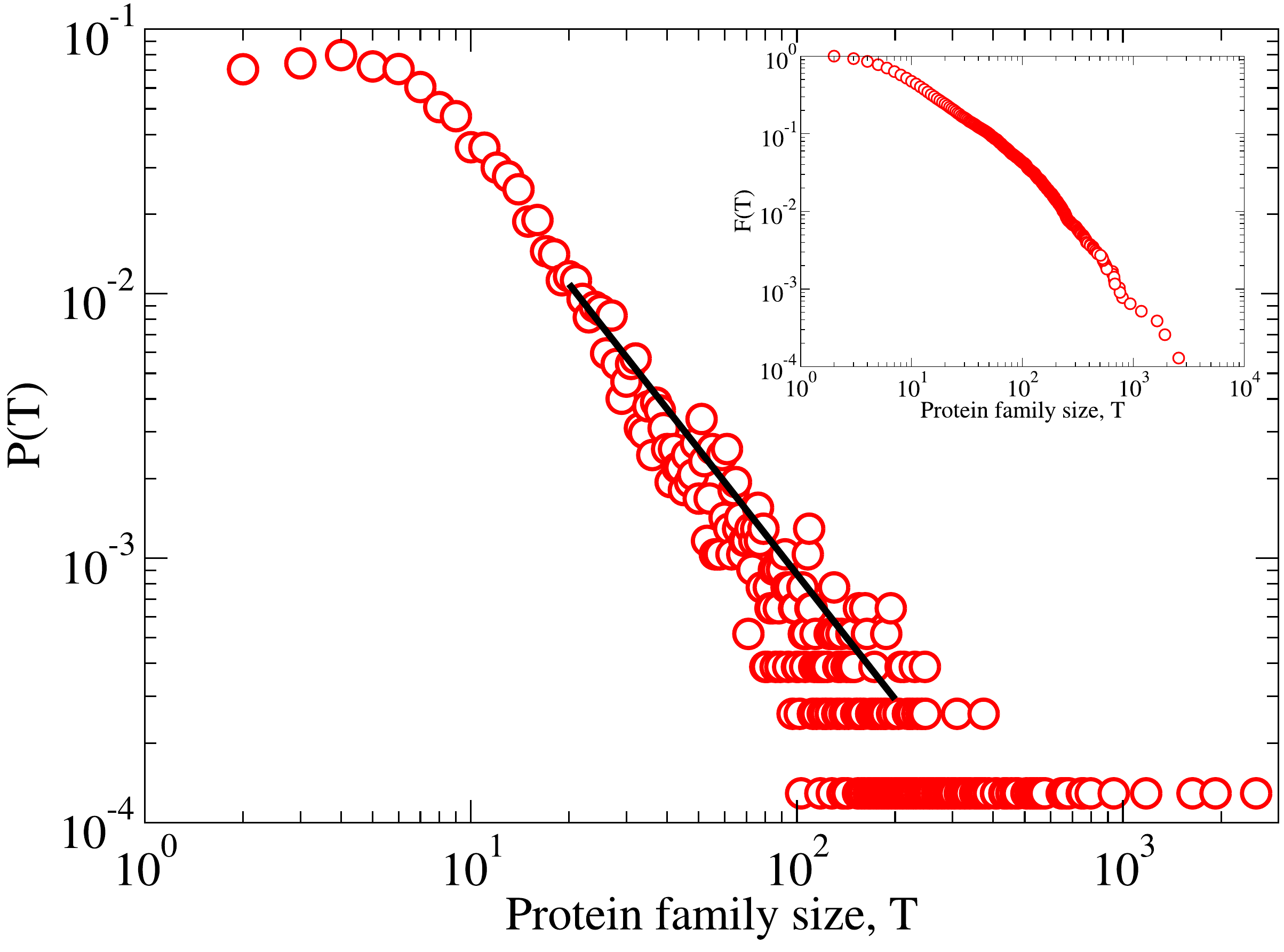}
 \caption{\textbf{Figure 1 - Protein family size distribution}}
 \label{fig1}
\end{figure}

\begin{figure}
 \centering
\includegraphics[width=1\textwidth]{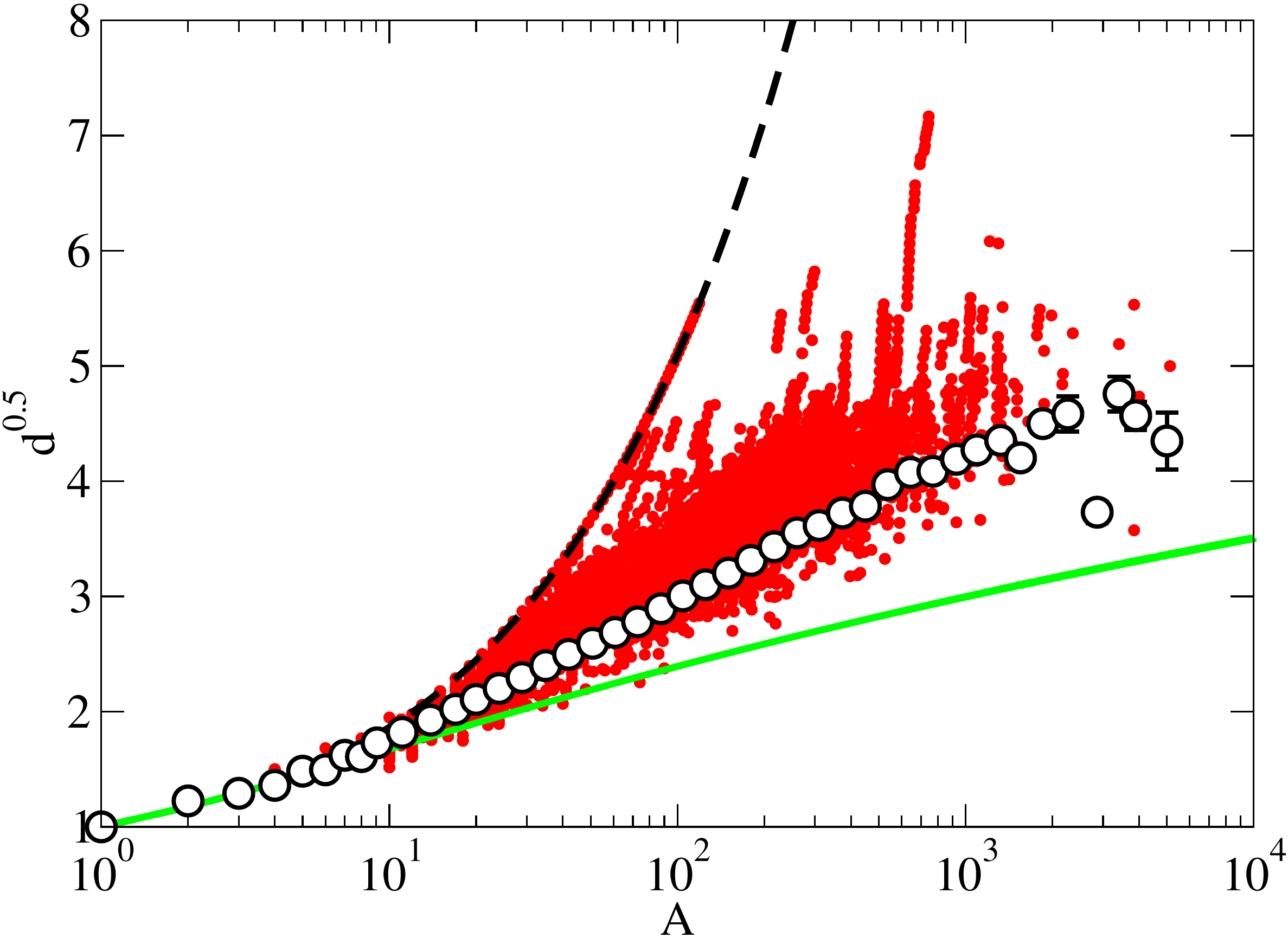}
 \caption{\textbf{Figure 2 - Depth scaling of protein phylogenies}}
 \label{fig2}
\end{figure}

\begin{figure}
 \centering
\includegraphics[width=1\textwidth]{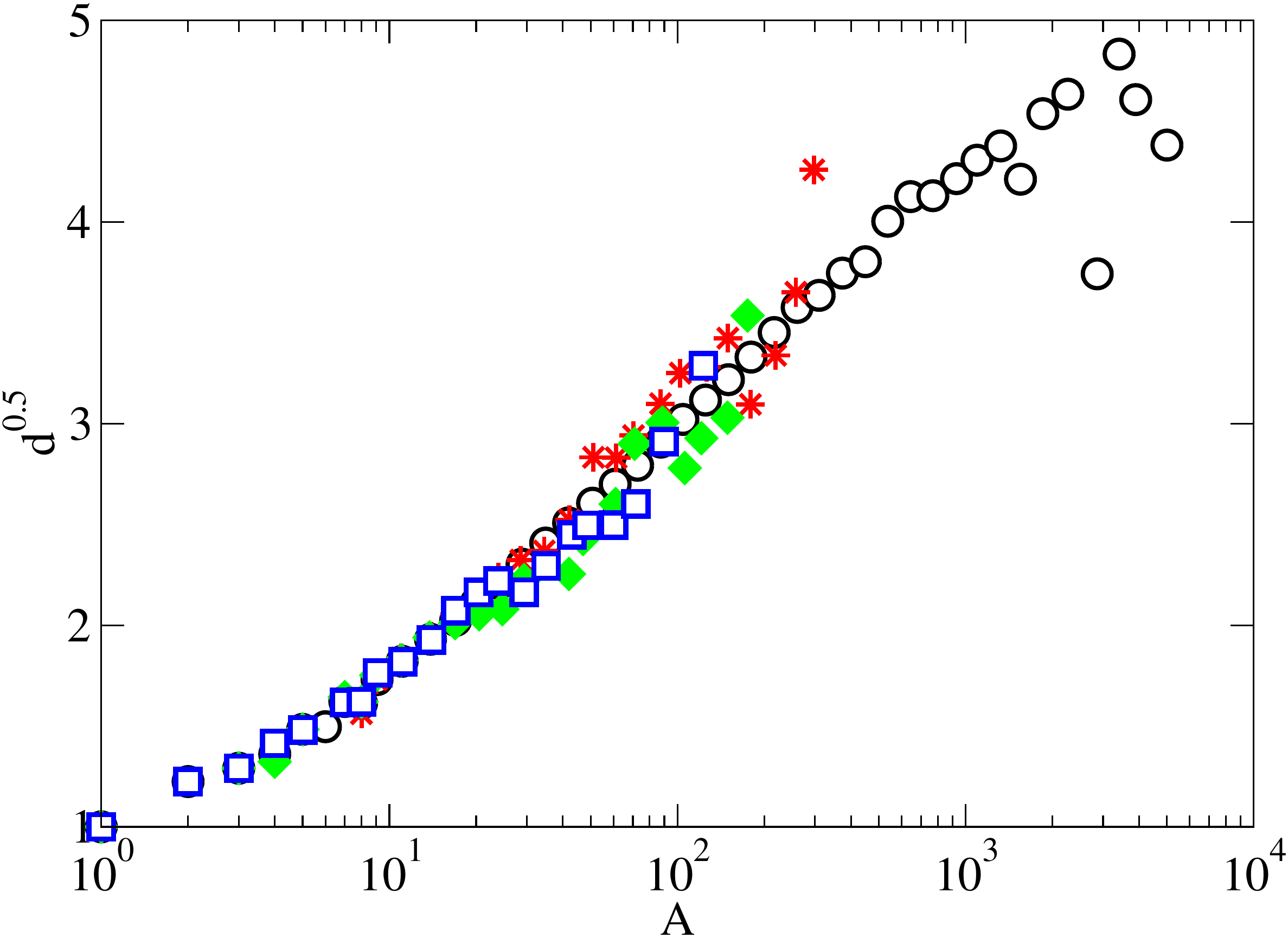}
 \caption{\textbf{Figure 3 - Depth scaling of different protein functions}}
 \label{fig3}
\end{figure}

\begin{figure}
 \centering
\includegraphics[width=1\textwidth]{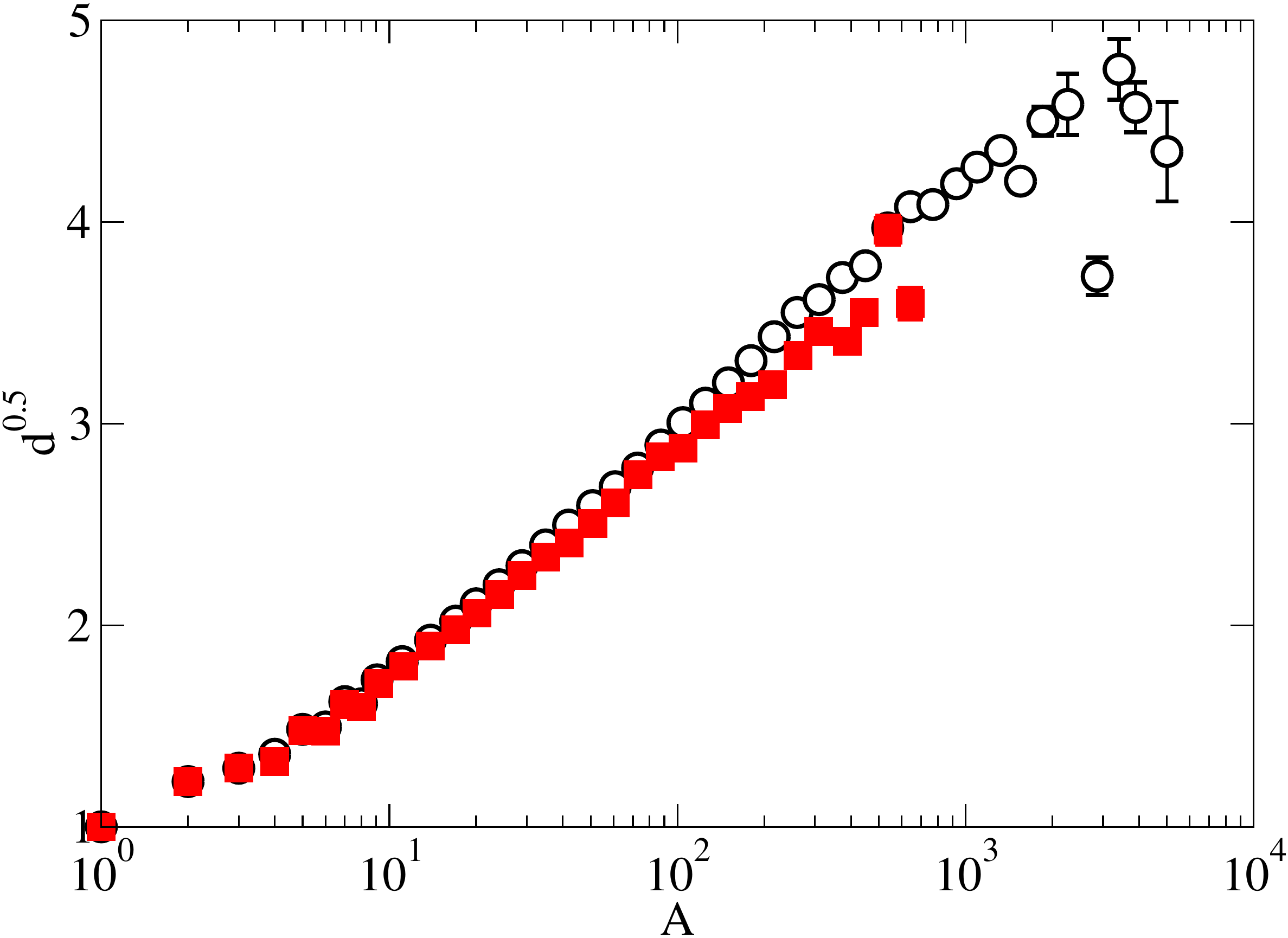}
 \caption{\textbf{Figure 4 - Protein vs organism phylogenies}}
 \label{fig4}
\end{figure}

\begin{figure}
 \centering
\includegraphics[width=0.5\textwidth]{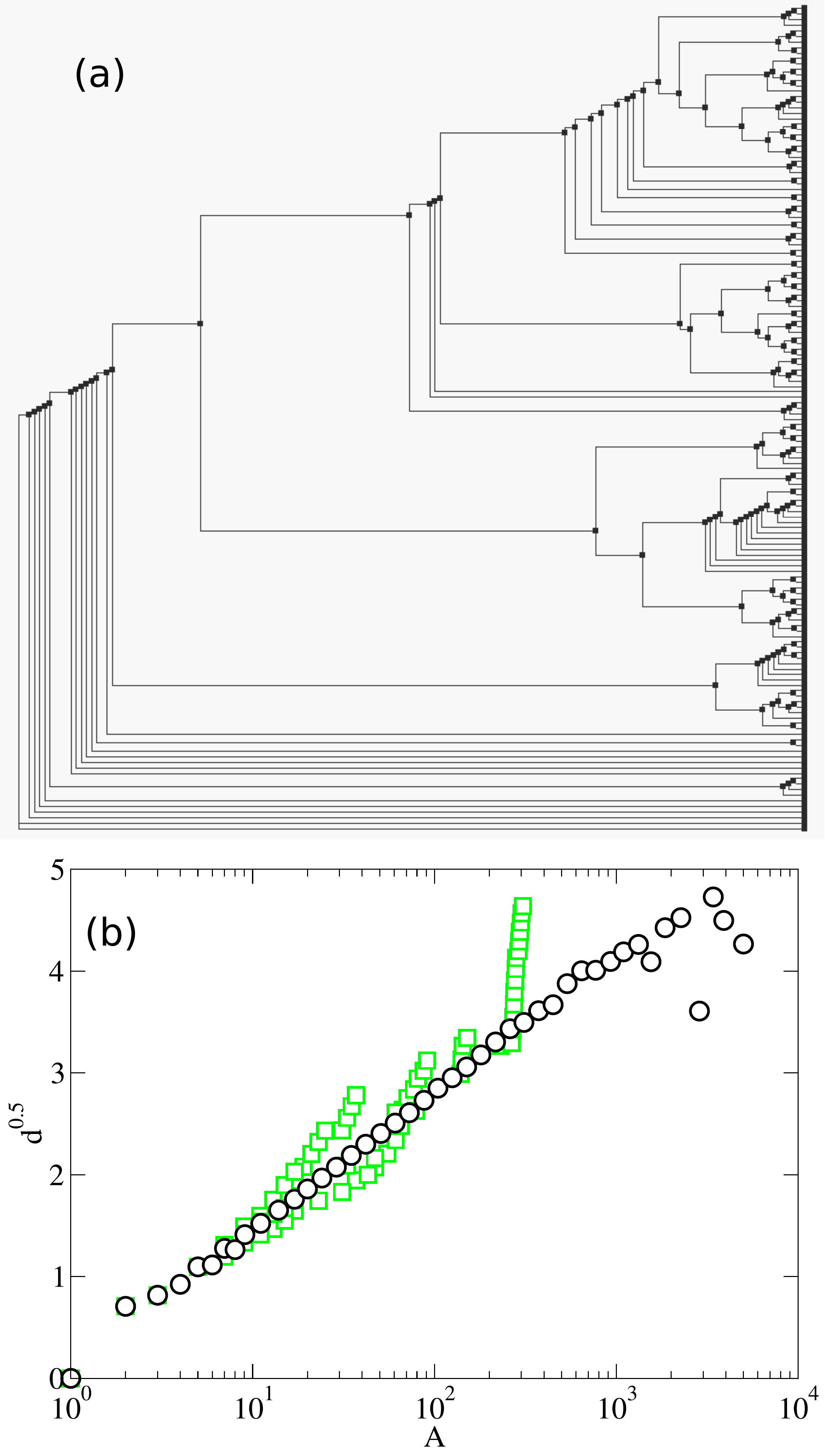}
 \caption{\textbf{Figure 5 - Mean depth behavior for a specific phylogenetic tree}}
 \label{fig5}
\end{figure}

\begin{figure}
 \centering
\includegraphics[width=1\textwidth]{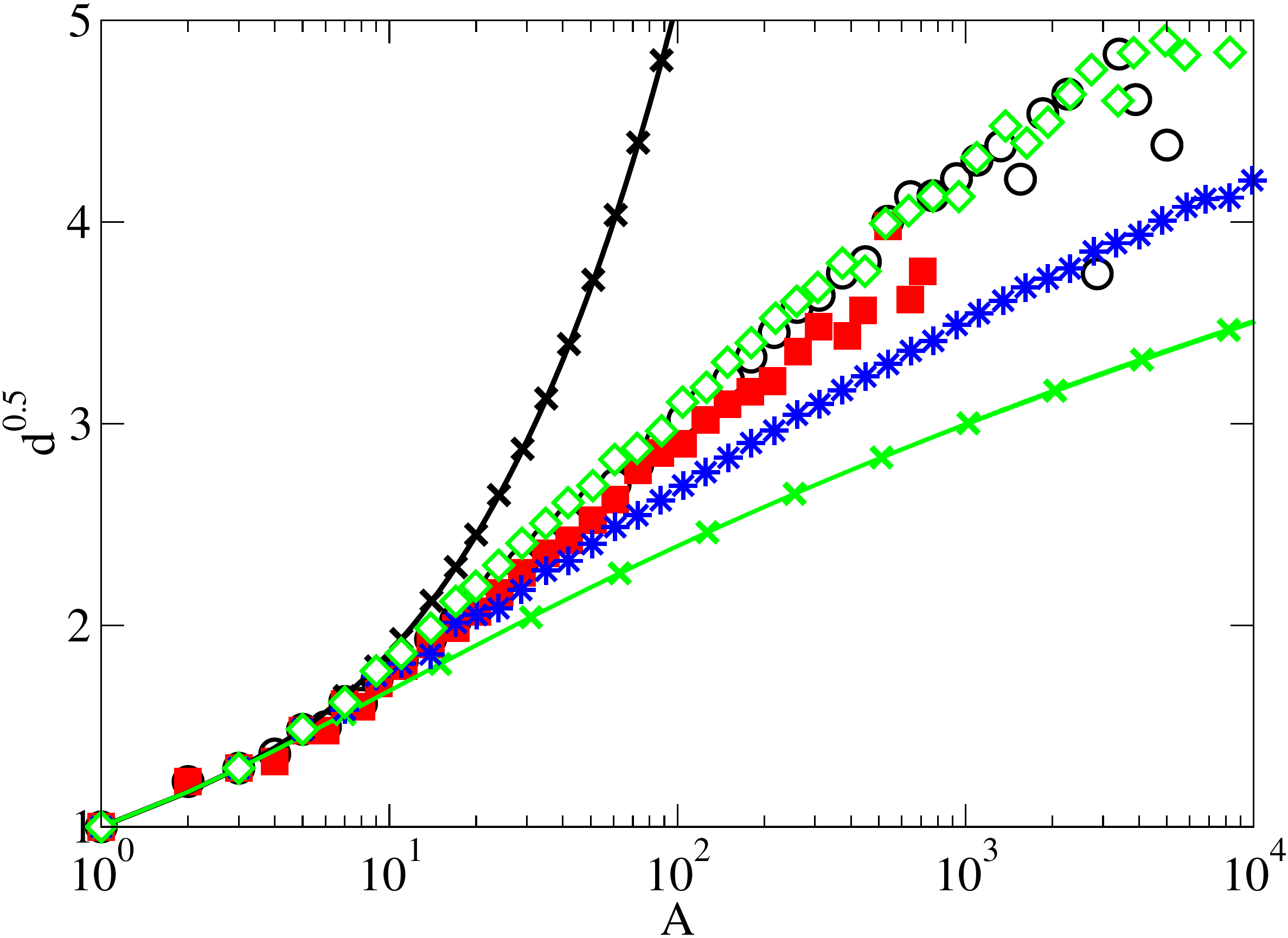}
 \caption{\textbf{Figure 6 - Depth scaling of the evolvability model}}
 \label{fig6}
\end{figure}

\begin{figure}
 \centering
\includegraphics[width=0.8\textwidth]{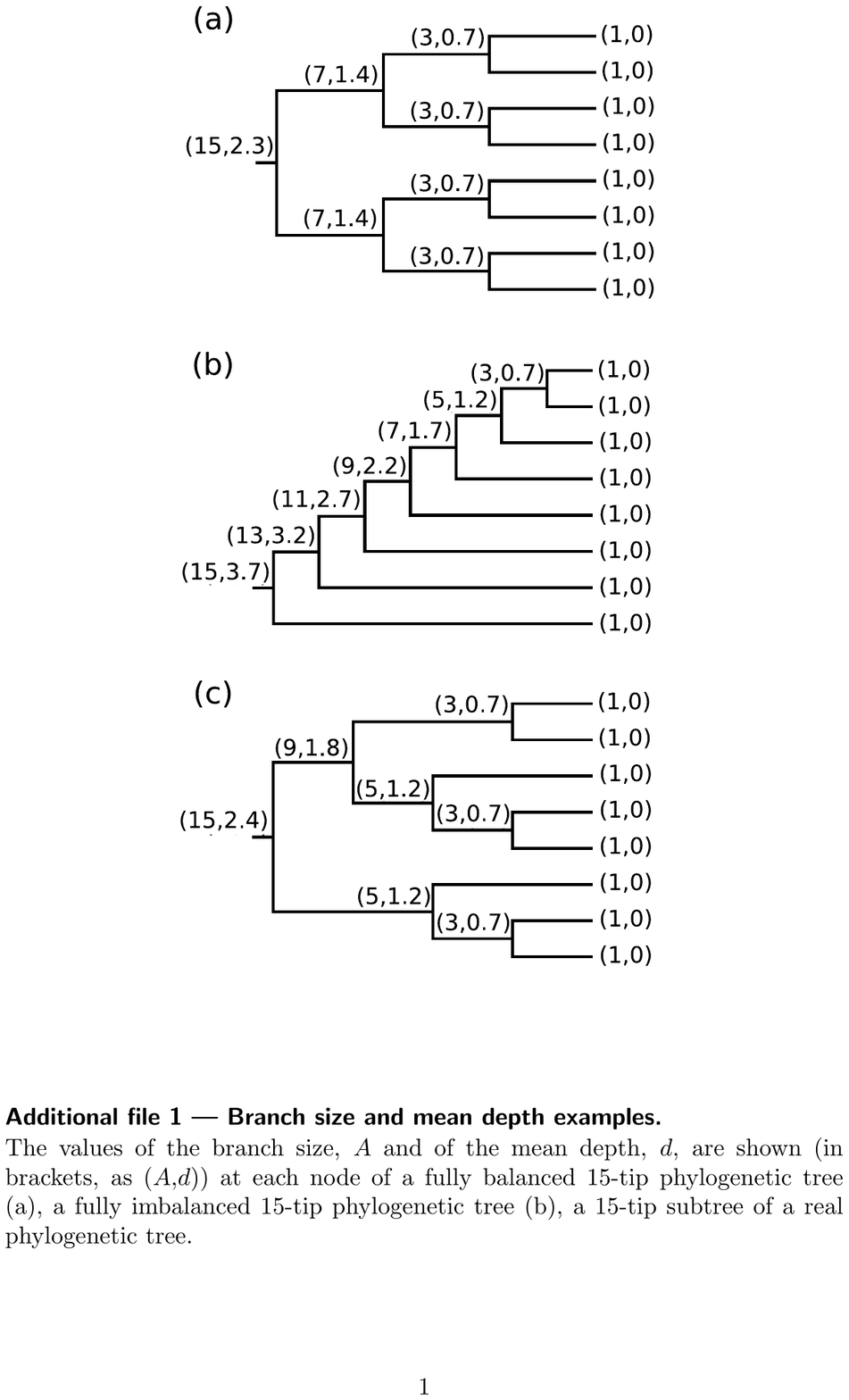}
 \caption{\textbf{Additional file 1 --- Branch size and mean depth examples}}
 \label{add1}
\end{figure}

\begin{figure}
 \centering
\includegraphics[width=0.8\textwidth]{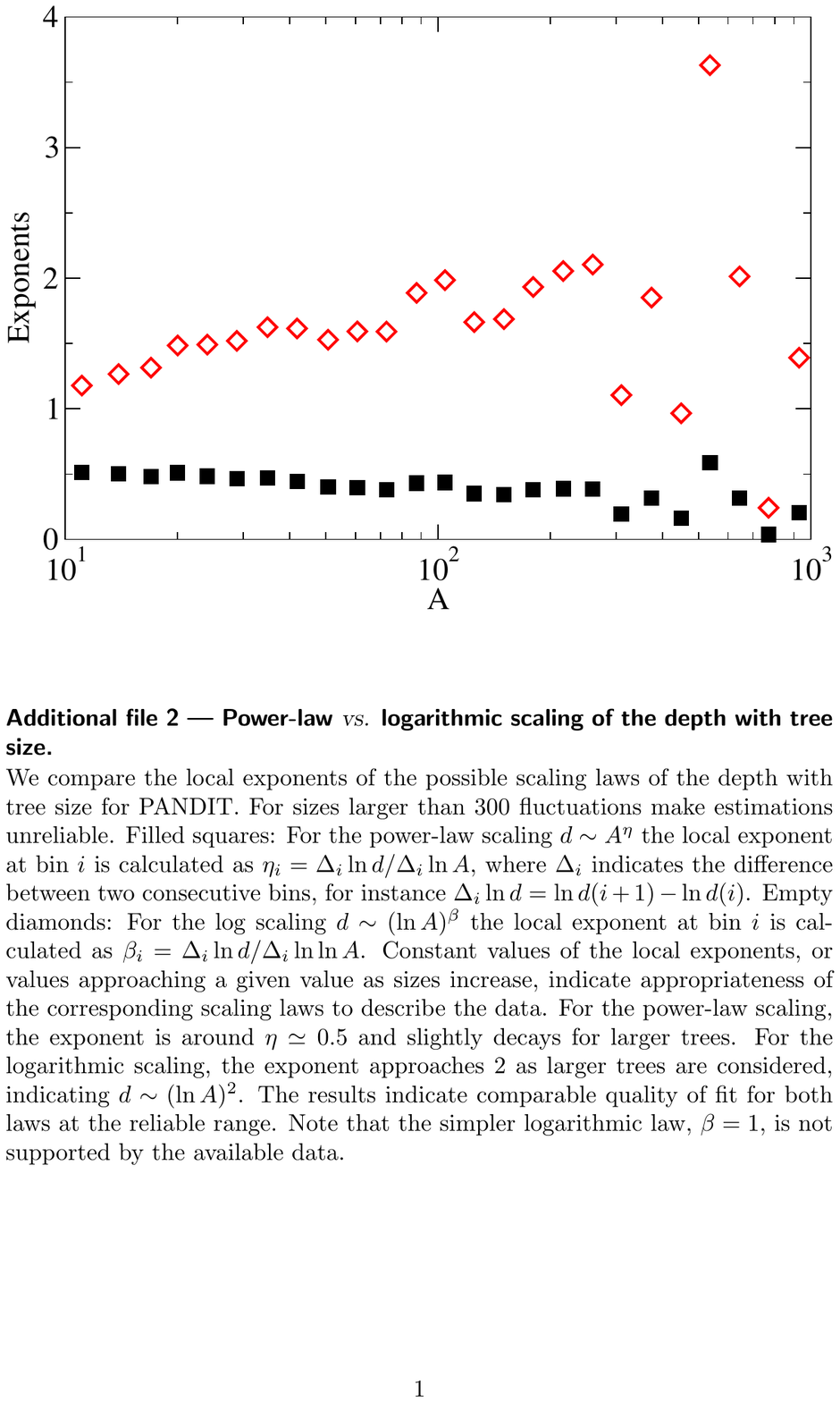}
 \caption{\textbf{Additional file 2 --- Power-law {\sl vs.} logarithmic scaling of the depth with tree size}}
 \label{add2}
\end{figure}

\begin{figure}
 \centering
\includegraphics[width=0.8\textwidth]{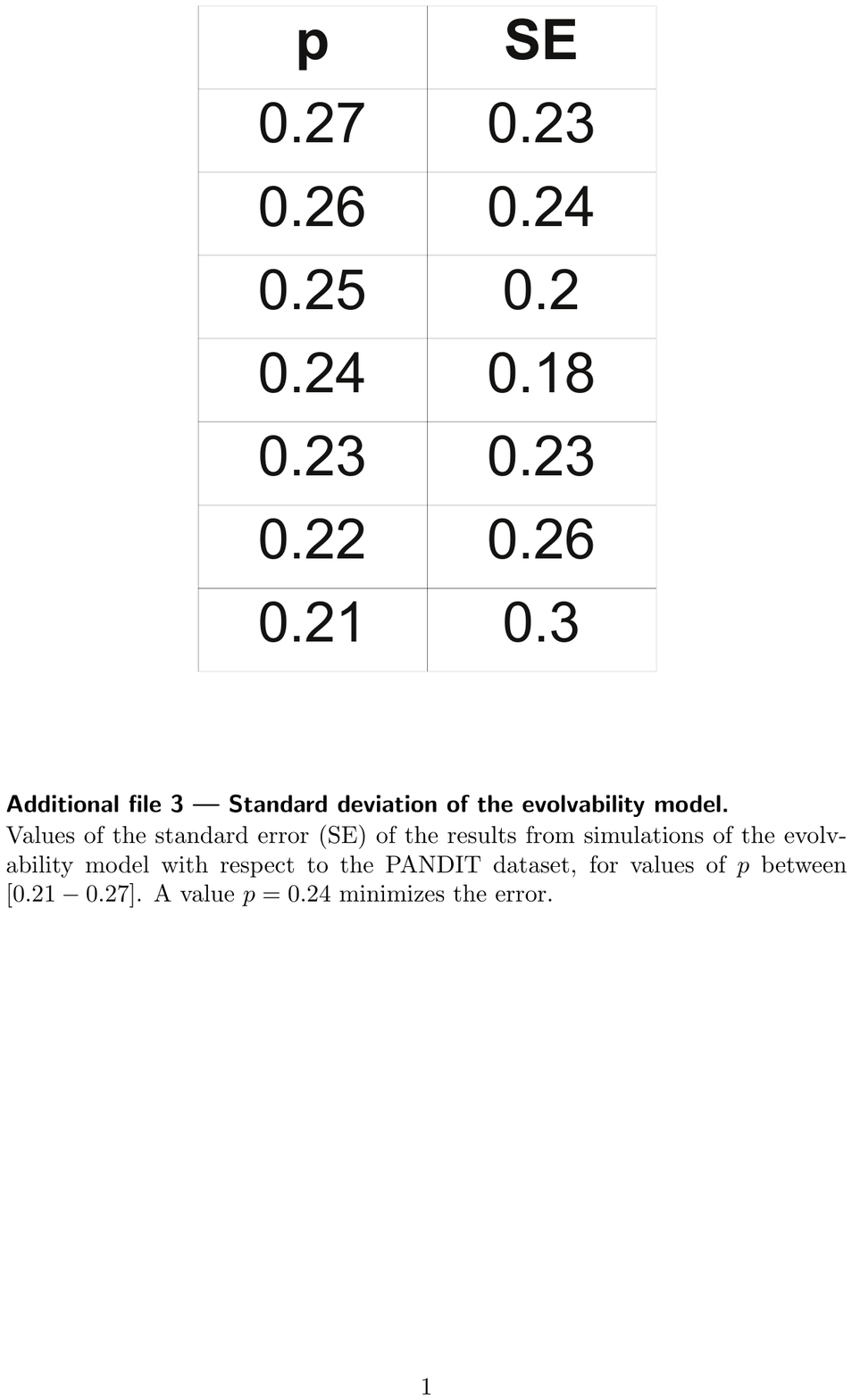}
 \caption{\textbf{Additional file 3 --- Standard deviation of the evolvability model}}
 \label{add3}
\end{figure}

\end{bmcformat}
\end{document}